\documentclass[prb,twocolumn,superscriptaddress]{revtex4-2}

\usepackage[utf8]{inputenc}
\usepackage{graphicx}

\usepackage{bm}
\usepackage{amssymb}
\usepackage{amsmath}
\usepackage{color}
\newcommand{\gf}[1]{\left\langle\!\!\!\left\langle #1 \right\rangle\!\!\!\right\rangle}

\newcommand{\kp}{k^{(+)}}

\newcommand{\fE}{\mathcal{E}}
\newcommand{\gpl}{\mathcal{G}}

\usepackage{xcolor}
\usepackage{color}
\usepackage{url}

\begin{document}

\title{Strongly coupled magnon-plasmon polaritons in graphene- 2D ferromagnet heterostructures}
\author{A. T. Costa}
\affiliation{International Iberian Nanotechnology Laboratory (INL), Av. Mestre José Veiga, 4715-330 Braga, Portugal}
\author{M. I. Vasilevskiy}
\affiliation{International Iberian Nanotechnology Laboratory (INL), Av. Mestre José Veiga, 4715-330 Braga, Portugal}
\affiliation{Department of Physics, Center of Physics, University of Minho, Campus of Gualtar, 4710-057, Braga, Portugal}
\author{J. Fern\'andez-Rossier}
\altaffiliation[On leave from ]{Departamento de F\'{\i}sica Aplicada, Universidad de Alicante, 03690,  Sant Vicent del Raspeig, Spain }
\affiliation{International Iberian Nanotechnology Laboratory (INL), Av. Mestre José Veiga, 4715-330 Braga, Portugal}
\author{N. M. R. Peres}
\affiliation{International Iberian Nanotechnology Laboratory (INL), Av. Mestre José Veiga, 4715-330 Braga, Portugal}
\affiliation{Department of Physics, Center of Physics, University of Minho, Campus of Gualtar, 4710-057, Braga, Portugal}
\date{\today}

\begin{abstract}
    Magnons and plasmons are two very different types of collective modes, acting on the spin and charge degrees of freedom, respectively.  At first sight, the formation of  hybrid plasmon-magnon polaritons in heterostructures of plasmonic and magnetic systems  would face two challenges, the small mutual interaction, via Zeeman coupling of the electromagnetic field of the plasmon with the spins,  and the energy mismatch, as  in most systems plasmons have energies in the eV range, orders of magnitude larger than magnons. Here we show that  graphene plasmons form polaritons with the magnons of  two-dimensional ferrromagnetic insulators, placed up to to half a micron apart, with Rabi couplings in the range of 100 GHz (dramatically larger than cavity QED magnonics).   This strong coupling is facilitated both by the small energy of graphene plasmons and the cooperative super-radiant nature of the plasmon-magnon coupling afforded by phase matching. We show that the Rabi  coupling can be  modulated both electrically and mechanically and we propose a attenuated total internal reflection experiment to implement ferromagnetic resonance experiments on 2D ferromagnets driven by plasmon excitation. 
\end{abstract}

\maketitle

%\section{Introduction}
Magnons are the elementary excitations of every magnetically ordered system, governing their 
low energy properties. Magnons attract renewed interest for several reasons. 
They can transport spin currents for applications in non-dissipative spintronics,\cite{cornelissen15} host topological order with chiral edge states,\cite{Owerre2016,mcclarty22} form exotic collective states such as Bose Condensates and spin superfluids,\cite{bunkov2010} and, most important for the scope of this work, they can couple to photons.\cite{yuan22,Soljacic2019,Dirnberger2022}
 
Magnons play a particularly important role in 2D magnets as their uncontrolled thermal 
proliferation \cite{mermin66} prevents long-range order. Thus, most prominent examples of
2D ferromagnets, such as VI$_3$,\cite{VI3_Raman} CrI$_3$\cite{CrI3neutron2018} and Fe$_3$GeTe$_2$, have a sizable gap in the magnon energy spectrum.
Experimental techniques that are very successful in producing and probing magnons in bulk 
ferromagnets are not easily adaptable to 2D systems due to the intrinsically small sample volume. 
For instance, the sensitivity of ferromagnetic
resonance is limited by the ratio between sample and detector sizes. Recent proposals, such as
ferromagnetic resonance force spectroscopy,\cite{PhysRevB.78.144410} address the challenge of
probing submicron-size samples, but are a long way from monolayer van der Waals magnets.
Cavity magnonics\cite{CavityMagnonicsBlanter2021} has also emerged as a way of enhancing
the coupling between exciting/probing fields and the magnetic sample. 
Rabi splittings of the order of 100~MHz have been obtained for micron-sized spheres
on resonant microwave cavities.~\cite{PhysRevLett.113.083603} Further enhancement 
in coupling strength, leading to Rabi splittings of a few GHz, has been achieved for 
macroscopic-sized ferromagnets in optical\cite{Flower_2019} and superconducting cavities.\cite{Golovchanskiy2021}

In this context, exciting and probing magnons efficiently in 2D ferromagnets remains a challenge. 
There are three main bottlenecks for the existing techniques. One is having 
a driving field of the right frequency: magnons in 2DFM have frequencies in the range
$\sim 0.25 - 1$~THz, whereas the highest frequencies achieved in FMR experiments
are $\sim 700$~GHz.\cite{HFFMR1999} This stems from a combination of the scarcity of microwave sources
of higher frequencies and the need to match the resonance frequency of a 
cavity. This brings forward the second challenge, the strength of the photon-magnon coupling.
The interaction of the magnetic field of light with matter is notoriously much weaker than that of the electric field. Placing the ferromagnetic sample in a 
resonant cavity enhances the coupling between the magnon and the cavity modes.
The frequencies of those modes, however, decrease as the cavity volume increases,
whereas the enhancement factor goes in the opposite direction. There is, thus, a
compromise between enhancement factor and resonance frequency that limits the sensitivity
of setups of this kind. This links to the third challenge, which is detector sensitivity.
Again, this is limited by the smallness of light's coupling to magnetic dipoles, and puts a constraint
on the minimum enhancement factor needed.

In regard to the frequency of the driving field, graphene plasmons come to
mind as prime candidates. Their frequencies can be tuned essentially continuously,
by gating graphene away from charge neutrality. Current experimental limits on such control
set the spectral range of graphene plasmons to a few THz within the wavelength range of 
interest to us. Graphene plasmons have been shown to form various kinds of polaritons in
van der Waals heterostructures.\cite{GarciadeAbajo2022}
Coupling to graphene plasmons has been proposed recently as a way to probe collective 
excitations in superconductor surfaces,~\cite{Costa2021PNAS} 2D superconductors,~\cite{Costa2021JPCM}
and excitons in insulators.~\cite{Nerl2017} The common theme of those works is the coupling between the 
strongly confined electric field associated with the graphene plasmon and the charges of the electrons in
the nearby system. The coupling to spins is more subtle, since it relies on the much smaller
magnetic-dipolar nature. It is known, however, that momentum and frequency matching can enhance
dramatically the coupling between light and an ensemble of quantum objects.\cite{Dicke1954}
With this in mind, we have studied the coupling between graphene plasmons and 2D magnons in a
van der Waals heterostructure.

Long range ferromagnetic order in 2D is only possible in the presence of magnetic anisotropy, on account of the
Mermin-Wagner theorem.~\cite{mermin66,Halperin2019} Spin-orbit coupling breaks
spin rotation symmetry, stabilizes long range magnetic order and opens-up a gap in the magnon spectrum at zero wave-vector, $q=0$. In many cases of interest, the magnon gap  in 2D ferromagnets is much larger than typical 
values in 3D. For instance,  the magnon gap of   CrI$_3$ monolayers, one of the most prominent 
2D magnents, has been reported to be in the 0.3-1.0~meV range.\cite{FMR_CrI3PRL2019,CrI3neutron2018}
For some materials this value can exceed 5~meV,~\cite{PtCl32019,Jiang2021}
putting the lowest energy magnon in the terahertz region. On the other hand,
the energy and wave vector of graphene plasmons may be tuned to match those of magnons in a 2DFM by adjusting the charge density of the graphene sheet. 
Thus, van der Waals heterostructures composed of 2DFM and plasmonic materials, such as graphene, may provide a platform to bridge the
terahertz gap in optoelectronics. Previous attempts in this direction have been aimed at the coupling between light and the orbital magnetic moments of electrons in conducting materials,\cite{Pendry2004} but here we focus on the spin magnetic moment, which is associated with
quantum magnetism.

\begin{figure}
    \centering
    \includegraphics[width=\columnwidth]{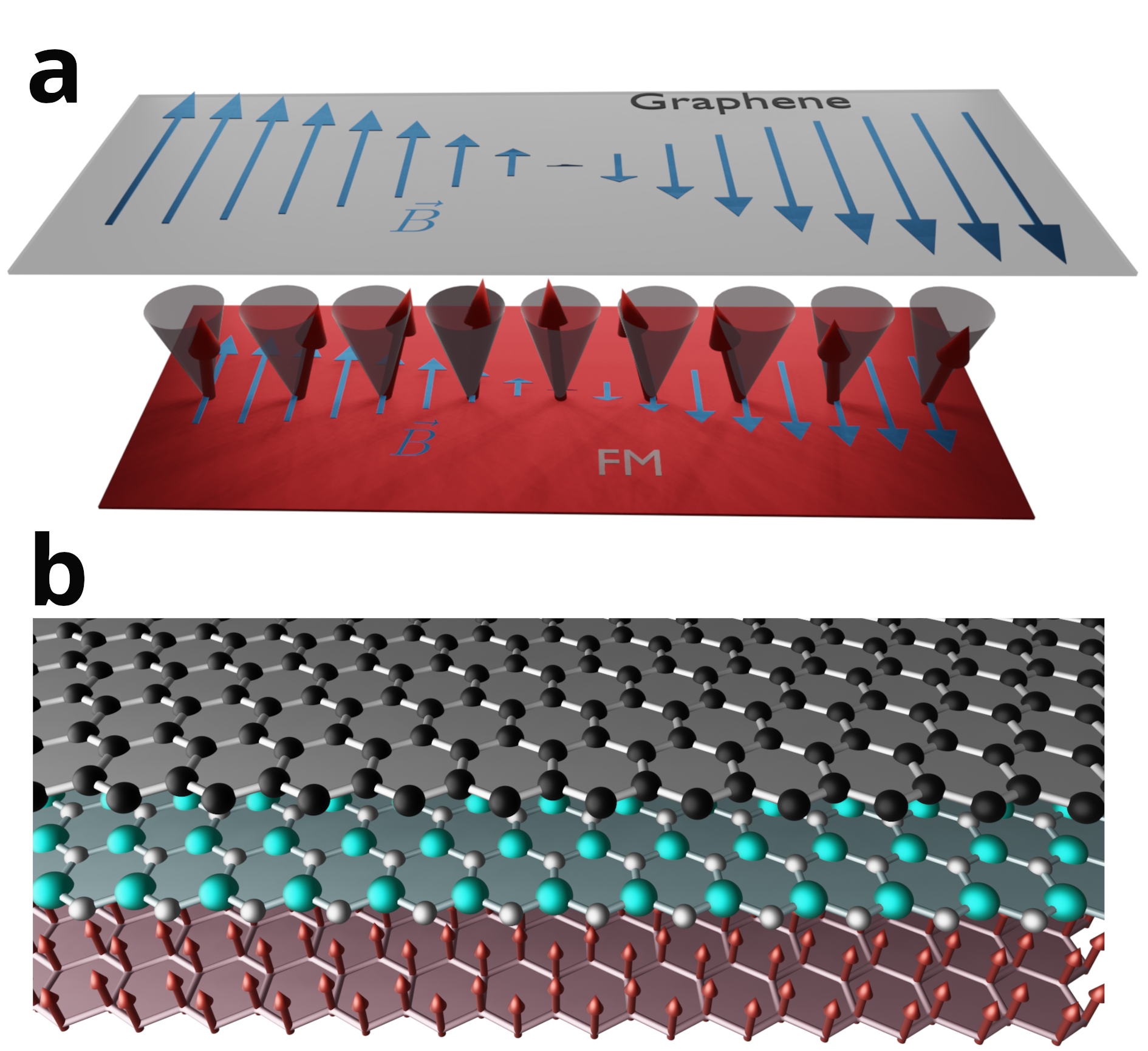}
    \caption{Schematic depiction of the  heterostructure where strong plasmon-magnon coupling
    is predicted to occur. a) Artistic rendition of the plasmon magnetic field, that emanates from the graphene layer and reaches  the magnetic layer, and the precession of the spins in a
    magnon state, with the same wave vector than the plasmon, in the magnetic layer. 
    b) Scheme of the structure that would display the effect, including a graphene monolayer, a boron nitride decoupling layer and the magnetic monolayer.  The plasmon-magnon coupling is large for decoupling layers as thick as 5$\mu$m.} 
    \label{fig:scheme}
\end{figure}

%\section{Model and Methods}

We consider a van der Waals heterostructure, depicted schematically in figure~\ref{fig:scheme}, 
composed of a 2D ferromagnet with off-plane easy axis and a graphene sheet, separated by a dielectric,
such as hexagonal boron-nitride, of thickness $z$ 
and relative dielectric constant $\epsilon$.

The 2D ferromagnet is described with a  spin Hamiltonian in the linear spin wave approximation.\cite{lado2017} 
The in-plane magnetic field of the plasmon is coupled to the local spins of the ferromagnet via Zeeman interaction,
\begin{equation}
    H_\mathrm{Z}=\mu_B\sum_l\hat{\vec{B}}(\vec{R}_l,z)\cdot \hat{\vec{\sigma}}_l ,
\end{equation}
where $\vec{R}_l$ is the 2D vector marking the position of unit cell $l$ in the 2D ferromagnet, $z$ is the vertical 
distance between the graphene sheet and the 2D ferromagnet, and $\vec{\sigma}$ are the dimensionless Pauli spin matrices that relate to the spin angular momentum through $\vec{S}=\frac{\hbar}{2} \vec{\sigma}$.

Using
the expression for the quantized field of the graphene plasmon given in ref.~\onlinecite{Henriques2021},
the Zeeman interaction with the TM plasmon magnetic field reads,
\begin{equation}
\begin{split}
    H_\mathrm{Z}=
    \mu_B
    \sum_l\sum_{\vec{q}}iF(q,z)\left[
    (q_y\hat{\sigma}^x_l - q_x\hat{\sigma}^y_l)e^{i\vec{q}\cdot\vec{R}_l}a_{\vec{q}} - \right. \\
 \left. (q_y\hat{\sigma}^x_l - q_x\hat{\sigma}^y_l)e^{-i\vec{q}\cdot\vec{R}_l}a^\dagger_{\vec{q}}  \right],
\end{split}
    \label{eq:Zeeman}
\end{equation}
where $a^\dagger_{\vec{q}}$ is the creation operator for a plasmon with wave vector $\vec{q}$ parallel 
to the graphene sheet. We note that the plasmon magnetic field lies in-plane, so that it generates a torque on the static magnetization. At the microscopic level, this entails the creation of magnons.
The coupling strength $F(q,z)$ is given by
\begin{equation}
    F(q,z) =
    -\epsilon\frac{\omega^2_\mathrm{pl}(q)}{c^2q\kappa_{\vec{q}}}\sqrt{\frac{\hbar}{2A\epsilon_0\omega_\mathrm{pl}(q)\Lambda(\vec{q})}}
    e^{-\kappa_{\vec{q}}|z|},
    \label{eq:Fqz}
\end{equation}
where $\hbar\omega_\mathrm{pl}(q)$ is the energy of a plasmon with wave vector $\vec{q}$, $\kappa_{\vec{q}}\equiv\sqrt{q^2-\epsilon\frac{\omega^2_{\vec{q}}}{c^2}}$, $\Lambda(q)$ is the
mode length of the plasmon (see supp. mat.), and $A$ is the area of the graphene sheet. The plasmon field decays exponentially, but
for the range of wave vectors relevant to this work the decay length is of the order of several microns,
thus presenting no practical concern.

We note that the coupling strength of a plasmon mode  with wave-vector $q$ with an atomic spin $\vec{S}_l$
is vanishingly small, as it scales with the inverse of $\sqrt{A}$. In contrast, for collective excitations 
such as magnons, it makes sense to transform the spins to a plane-wave basis,
\begin{equation}
    \hat{\vec{\sigma}}_{\vec{k}} \equiv \frac{1}{\sqrt{N}}\sum_l e^{i\vec{k}\cdot\vec{R}_l}\hat{\vec{\sigma}}_l ,
\end{equation}
where %$\mu\in\{x,y,z\}$,
$\vec{k}$ is a wave vector in the Brillouin zone of the 2DFM, and $N$ is the number of unit cells. After applying this transformation to Eq.~\ref{eq:Zeeman}
we obtain
\begin{equation}
\begin{split}
    H_\mathrm{Z}=
    \mu_B\sqrt{N}\sum_l\sum_{\vec{q}}iF(q,z)\left[
    (q_y\hat{\sigma}^x_{\vec{q}} - q_x\hat{\sigma}^y_{\vec{q}})a_{\vec{q}} - \right. \\ \left. (q_y\hat{\sigma}^x_{-\vec{q}} - q_x\hat{\sigma}^y_{-\vec{q}})a^\dagger_{\vec{q}}  \right].
\end{split}
    \label{eq:Zeeman2}
\end{equation}
Compared to the case of atomic spins, the magnon-plasmon coupling is enhanced by a factor $\sqrt{N}$,
where $N$ is the number of spins, resulting in a Rabi-coupling  that does not depend anymore on system size, 
as $N\propto A$. Thus, magnon-plasmon coupling is enhanced due to the phase-matching of the plasmon field 
to a macroscopic number of phase-locked precessing spins. 
%Here, as we show below,
%the requirements of energy and momentum conservation ensure that the coupled plasmon and magnon 
%are phase-matched, leading to the coupling enhancement.

The quantized Hamiltonian for   plasmons in graphene  reads:
\begin{equation}
H_\mathrm{plasmon}   \equiv \sum_{\vec{q}}\hbar\omega_\mathrm{pl}(q)a^\dagger_{\vec{q}}a_{\vec{q}}.
\end{equation}
where their energy dispersion curve is given by~\cite{NunosBook}
\begin{equation}
    \hbar\omega_\mathrm{pl}(q) = \sqrt{ \frac{2\alpha E_F}{\epsilon}\left[\sqrt{(\alpha E_F)^2 + (\hbar c q)^2} - \alpha E_F\right] },
\end{equation}
Here, $E_F$ is graphene's Fermi energy, $\epsilon$ is the average dielectric constant of the two media surrounding
the graphene sheet, $\alpha$ is the fine structure constant, $c$ is the speed of light and $q$ is the
plasmon's (in-plane) wave vector.

To study the effect of plasmon-magnon coupling we adopt a description of magnons in terms
of linearized Holstein-Primakoff bosons,\cite{holstein40}
\begin{equation}
    \hat{\sigma}^-_{l}\simeq\sqrt{2S}b_{l}^\dagger,\,\,\, \hat{\sigma}^+_{l}\simeq\sqrt{2S}b_{l},
\end{equation}
where $\hat{\sigma}^{+,-}$ are the ladder operators acting on the spin located at site $l$; their
magnitude $S$ is assumed to be the same throughout the whole material. The operators $b^\dagger_{l}$ and $b_{l}$ 
respectively create and annihilate a localized spin flip excitation at site $l$. 
Assuming translation symmetry in the 2D ferromagnet we can rewrite the HP bosons in reciprocal space. Then, 
the Hamiltonian for bare magnons has the form
\begin{equation}
    H_\mathrm{m} = \sum_{\vec{k}}\hbar\omega_{\rm mag}(\vec{k})b^\dagger_{\vec{k}}b_{\vec{k}}.
\end{equation}
The wave vectors $\vec{k}$ span the Brillouin zone of the 2D ferromagnet. The
function $\hbar\omega_{\rm mag} (\vec{k})$ is the dispersion relation for the bare magnons. 
For small momenta, we have  $\hbar\omega\simeq \hbar\omega_{0}+ \rho k^2$, where the first term 
is the magnon gap and the second provides the dispersion due to the exchange-driven spin stiffness $\rho$.

 For plasmons with energies $\hbar\omega_\mathrm{pl}\sim 1$~meV
(thus close to that of uniform magnons in typical 2DFM), and typical graphene doping levels ($E_F\sim 100$~meV), $q\lesssim 0.1\mu\mathrm{m}^{-1}$. 
This is tiny compared to the linear dimensions of the magnon Brillouin zone ($\sim 10^4\mathrm{\mu\mathrm{m}}^{-1}$), so that the dispersion of the magnon states
is negligible in that wave-vector window.

 After 
transforming the Zeeman Hamiltonian to the HP representation in reciprocal space it reads
\begin{eqnarray}
    H_\mathrm{Z}=
    \sum_{\vec{k}}
    %\left\{
    \left[ \hbar\Omega_{\vec{k}}(z)b^\dagger_{\vec{k}} + \hbar\Omega^*_{-\vec{k}}(z) b_{-\vec{k}} \right](a_{\vec{k}}+  a^\dagger_{\vec{k}} )
    %\right.
    %\nonumber\\ 
    %\left. \left[ \Omega(-\vec{k},z)b^\dagger_{-\vec{k}} +\Omega^*(\vec{k},z) b_{\vec{k}} \right] a^\dagger_{\vec{k}}  \right\}. \nonumber\\
\end{eqnarray}
The coupling strength  is given by
\begin{equation}
    \hbar\Omega_{\vec{k}}(z)\equiv \mu_B\sqrt{2NS}F(k,z)\kp ,
    \label{eq:couplingzeeman}
\end{equation}
where $z$ is the distance between the graphene sheet and the 2DFM, $N$ is the number of spins in the 2DFM, 
and $\kp\equiv k_x+ik_y$. The function $F(k,z)$ has been defined in Eq.~\ref{eq:Fqz}. Notice that the plasmon-magnon
coupling is diagonal in wave vector, meaning that each bare plasmon of wave vector $\vec{k}$ couples only to
magnons with the same wave vector. 

\begin{figure}
    \includegraphics[width=\columnwidth]{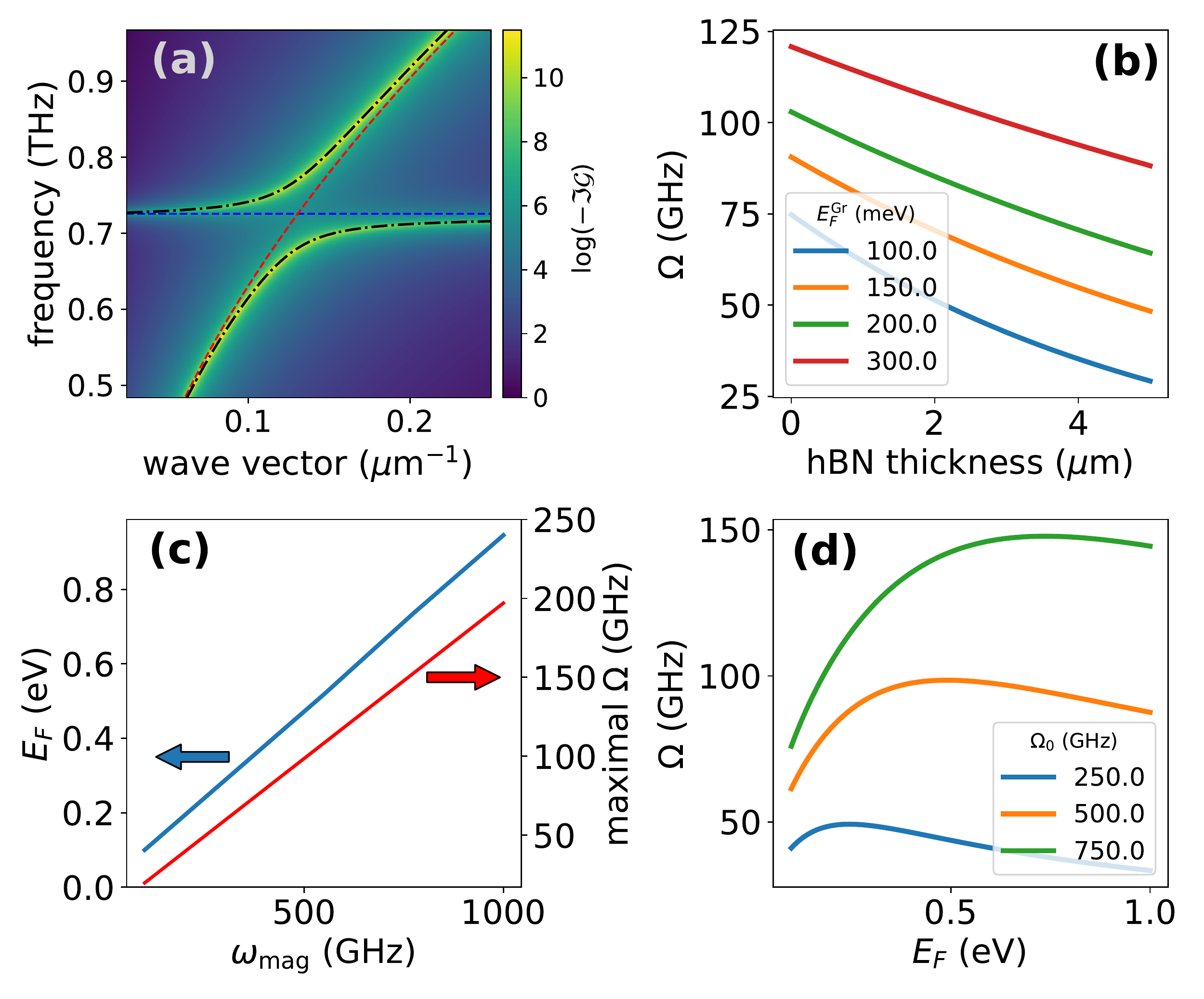}
    \caption{Main features of the hybrid plasmon-magnons excitation. a) Spectral density as a function of frequency and wave vector for fixed graphene doping ($E_F=200$~meV) and
    hBN thickness (10 nm). The magnon gap has been set at 3~meV, corresponding to a frequency of $\sim$ 0.73~THz. The dashed blue 
    and red lines correspond to the dispersion relations of the bare magnon and plasmon, respectively. The black dot-dashed lines are the approximate dispersions of the hybrid plasmon-magnon modes given by Eq.~\ref{eq:analyticdispersions}. b) Rabi splitting as a function
    of hBN thickness for different graphene doping levels. The magnon gap is the same as in a). Panel c) shows the Fermi energy of graphene for which the maximum Rabi splitting is obtained, as a function of the magnon frequency (blue curve, left vertical axes), and
    the respective maximal splitting (red curve, right vertical axis) . Panel d) shows Rabi splitting as a function of graphene gating level for different magnon frequencies, at a fixed hBN thickness of 10~nm.}
    \label{fig:specdens1}
\end{figure}

If the terms proportional to $b_{-\vec{k}}a_{\vec{k}}$ and $b^\dagger_{\vec{k}}a^\dagger_{\vec{k}}$
in Eq.~\ref{eq:couplingzeeman} are neglected, the remaining Hamiltonian can be mapped onto a single-particle
problem, leading to approximate analytic forms for the dispersion relations of the two hybrid
plasmon-magnon modes,
\begin{equation}
    E_{\pm} = %\frac{\hbar}{2}\left[\omega_{pl}(k) + \omega_{\rm mag}(k) \pm \sqrt{\left(\omega_{pl}-\omega_{\rm mag}(k)\right)^2+4|\frac{\Omega}{\hbar}(k,z)|^2}\right] 
     \hbar\omega_{+} \pm \sqrt{(\hbar\omega_{-})^2+|\hbar\Omega_{\vec{k}}(z)|^2}
    .
    \label{eq:analyticdispersions}
\end{equation}
where $\omega_{\pm}=\frac{\omega_\mathrm{pl}\pm\omega_\mathrm{mag}}{2}$. 
This equation predicts  a gap opening of magnitude $\hbar\Omega_{\vec{k}}(z)$ at the crossing 
frequency where the plasmon-magnon detuning $\omega_{-}$ vanishes.

In the following we treat the complete magnon-plasmon Hamiltonian, including the non-conserving 
terms $ba$ and $a^{\dagger}b^{\dagger}$, by analyzing the plasmon Green function (see Methods),
%\begin{equation}
%    \gpl(\vec{k};t)\equiv -i\theta(t)\left\langle \left[a_{\vec{k}}(t),a^\dagger_{\vec{k}}\right]\right\rangle,
%\end{equation}
which can be probed in near-field optical experiments. Since the plasmon-magnon coupling is linear the equations
of motion for all Green functions can be solved analytically. Their explicit expressions are given in the
Methods section. Here we will highlight the most relevant features by plotting the plasmon spectral
density, $-{\rm Im}\gpl(\vec{k};\omega)$, that is of course affected by the coupling to magnons. 
%It is easy to show that, since the system is transitionally invariant,
%$ \gpl(\vec{k},\vec{k}';t)= \gpl(\vec{k};t)\delta_{\vec{k},\vec{k}'}$.

%%%%%%%%%%%%%%%%%%%%%%%%%%%%%%%%%%%%%%
%\section{Results and discussion}
%%%%%%%%%%%%%%%%%%%%%%%%%%%%%%%%%%%%%%

In Fig.~\ref{fig:specdens1}a show the spectral density for the case where magnon gap 
 has $\omega_{\rm mag}(0)= $3~meV, corresponding to a 2DFM such as Fe$_3$GeTe$_2$.~\cite{FeGeTe_neutron,FeGeTeMAE2020} 
 In that figure we also show the dispersion curves for the bare plasmon and magnon.  The formation of a plasmon-magnon polariton with a Rabi splitting larger than 100 GHZ, dramatically larger than the values reported in cavity magnonics\cite{CavityMagnonicsBlanter2021}  is apparent. 

Interestingly, the magnitude of the Rabi coupling $\Omega$ can be tuned mechanically, by controlling the graphene-ferromagnet distance $z$, as we show in Fig.~\ref{fig:specdens1}b. In this energy range the plasmon decaying rate in 
the direction perpendicular to the graphene layer is small, which means that the plasmon-magnon coupling is sizeable even 
for graphene-2DFM distances of the order of 1~$\mu$m, where interlayer exchange is completely negligible.

The Rabi coupling can be further tuned electrically, controlling the graphene Fermi energy $E_F$ with a back gate, 
as we show in Fig.~\ref{fig:specdens1}b,c for three different 2D ferromagnets. For a given magnon energy, there is an optimal value of $E_F$ that maximizes the Rabi coupling strenght, as we show Fig.~\ref{fig:specdens1}d. We thus see that in a wide range of experimentally relevant parameters, the intrinsic magnon-plasmon Rabi coupling due to Zeeman coupling can be in the larger than 50~GHz. The estimated Rabi coupling is a lower bound, coming from the intrinsic Zeeman interaction, and  additional contributions to the Rabi coupling can occur when the magnon anisotropy  gap is sensitive to the plasmon electric field.

\begin{figure}
    \includegraphics[width=\columnwidth]{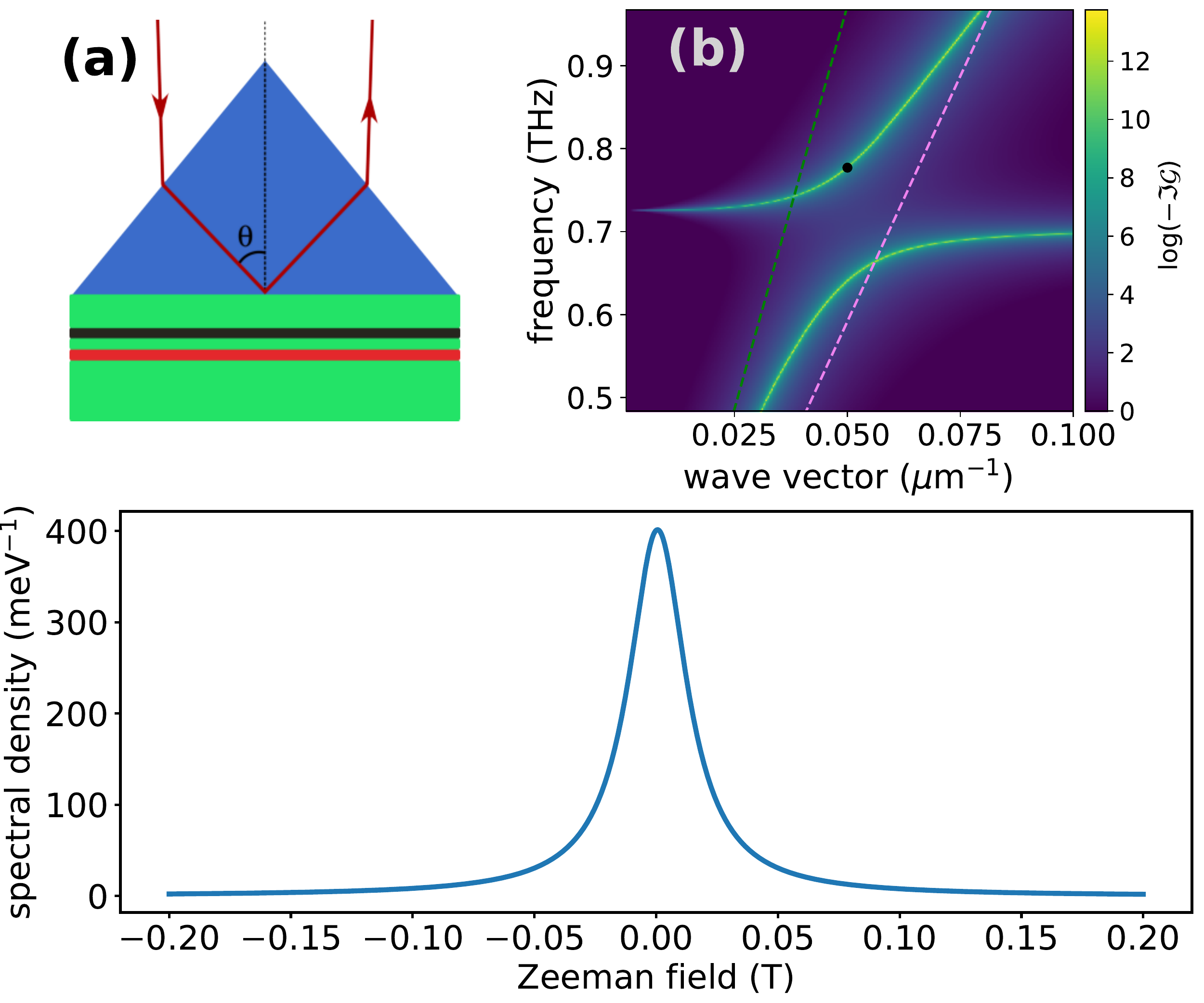}
    \caption{Attenuated total reflection experiment to probe the plasmon-magnon coupling. In a) we show
    a scheme of the setup. In b) we show the spectral density as a function of wave vector and frequency
    for a magnon energy of 1~meV and graphene doping corresponding to a Fermi energy of 100~meV.
    The spectral window probed by this experiment lies between the light dispersion relations 
    within hBN (green dashed line) and germanium (violet dashed line). c) Spectral density for
    the wave vector and frequency indicated by the black dot in panel b, as a function of and external
    magnetic field perpendicular to the plane of the heterostructure. The plasmon lifetime has been
    chosen as $\sim 5$~ns, in line with the intrinsic lifetimes given in Reference~\onlinecite{Principi2013}.}
    \label{fig:ATR}
\end{figure}

We now propose to take advantage of the magnon-plasmon coupling to carry out  ferromagnetic resonance of monolayers using an attenuated total reflection set-up (see Fig.~\ref{fig:ATR}a).
Exciting plasmons directly with optical beams is impossible due to the kinematic mismatch between plasmons and
propagating light.~\cite{NunosBook} By placing a prism of a high dielectric constant material on top of the hBN layer, 
it is possible to generate evanescent waves within the hBN that will excite the surface polaritons of the heterostructure. 
Whenever the in-plane component of the wave vector of light matches that of a polariton with the same frequency, there is a dip in the reflected intensity. The in-plane wave vector can be controlled via the incidence angle. With this set up, it is possible to excite polaritons whose wave
vectors and frequencies lie between the light cones inside hBN and the dielectric of which the prism is 
made. Germanium, for instance, would be a convenient material to use for the prism. It is transparent to electromagnetic
radiations of frequencies below 1~THz and its relative dielectric constant within the same frequency range is $\epsilon_\mathrm{Ge}\approx 16$.\footnote{ \protect{Other methods for exciting surface 
plasmon-polaritons are available, such as patterning gratings on top of graphene, which allow attaining larger wave vectors; here, however, we are interested in small wave vectors, for which the ATR is appropriate.} }

In Fig.~\ref{fig:ATR}a we plot the spectral density for 
a magnon gap of $\hbar\omega_0=1$~meV and graphene gating voltage corresponding to $E_F=100$~meV.
The dispersions for light inside hBN and germanium are plotted as dashed lines, to mark the spectral region probed by the
experiment. The black dot in Fig.~\ref{fig:ATR}b marks the point at which the plasmon-magnon spectral density
is probed. By applying an external magnetic field perpendicular to the structure, we shift the magnon energy, thereby changing
the spectral density and, consequently, the device's reflection coefficient. In Fig.~\ref{fig:ATR}c) we plot the spectral 
density, for fixed wave vector and frequency, as a function of the external magnetic field. The sharp peak heralds the
magnetic nature of the polariton being probed in this setup. This approach would permit to tackle the three issues that make FMR in 2D magnets challenging and could open a new venue to explore collective spin excitations in 2D systems.

\bibliography{references}

\begin{thebibliography}{34}%
\makeatletter
\providecommand \@ifxundefined [1]{%
 \@ifx{#1\undefined}
}%
\providecommand \@ifnum [1]{%
 \ifnum #1\expandafter \@firstoftwo
 \else \expandafter \@secondoftwo
 \fi
}%
\providecommand \@ifx [1]{%
 \ifx #1\expandafter \@firstoftwo
 \else \expandafter \@secondoftwo
 \fi
}%
\providecommand \natexlab [1]{#1}%
\providecommand \enquote  [1]{``#1''}%
\providecommand \bibnamefont  [1]{#1}%
\providecommand \bibfnamefont [1]{#1}%
\providecommand \citenamefont [1]{#1}%
\providecommand \href@noop [0]{\@secondoftwo}%
\providecommand \href [0]{\begingroup \@sanitize@url \@href}%
\providecommand \@href[1]{\@@startlink{#1}\@@href}%
\providecommand \@@href[1]{\endgroup#1\@@endlink}%
\providecommand \@sanitize@url [0]{\catcode `\\12\catcode `\$12\catcode
  `\&12\catcode `\#12\catcode `\^12\catcode `\_12\catcode `\%12\relax}%
\providecommand \@@startlink[1]{}%
\providecommand \@@endlink[0]{}%
\providecommand \url  [0]{\begingroup\@sanitize@url \@url }%
\providecommand \@url [1]{\endgroup\@href {#1}{\urlprefix }}%
\providecommand \urlprefix  [0]{URL }%
\providecommand \Eprint [0]{\href }%
\providecommand \doibase [0]{https://doi.org/}%
\providecommand \selectlanguage [0]{\@gobble}%
\providecommand \bibinfo  [0]{\@secondoftwo}%
\providecommand \bibfield  [0]{\@secondoftwo}%
\providecommand \translation [1]{[#1]}%
\providecommand \BibitemOpen [0]{}%
\providecommand \bibitemStop [0]{}%
\providecommand \bibitemNoStop [0]{.\EOS\space}%
\providecommand \EOS [0]{\spacefactor3000\relax}%
\providecommand \BibitemShut  [1]{\csname bibitem#1\endcsname}%
\let\auto@bib@innerbib\@empty
%</preamble>
\bibitem [{\citenamefont {Cornelissen}\ \emph {et~al.}(2015)\citenamefont
  {Cornelissen}, \citenamefont {Liu}, \citenamefont {Duine}, \citenamefont
  {Youssef},\ and\ \citenamefont {Van~Wees}}]{cornelissen15}%
  \BibitemOpen
  \bibfield  {author} {\bibinfo {author} {\bibfnamefont {L.}~\bibnamefont
  {Cornelissen}}, \bibinfo {author} {\bibfnamefont {J.}~\bibnamefont {Liu}},
  \bibinfo {author} {\bibfnamefont {R.}~\bibnamefont {Duine}}, \bibinfo
  {author} {\bibfnamefont {J.~B.}\ \bibnamefont {Youssef}},\ and\ \bibinfo
  {author} {\bibfnamefont {B.}~\bibnamefont {Van~Wees}},\ }\href@noop {}
  {\bibfield  {journal} {\bibinfo  {journal} {Nature Physics}\ }\textbf
  {\bibinfo {volume} {11}},\ \bibinfo {pages} {1022} (\bibinfo {year}
  {2015})}\BibitemShut {NoStop}%
\bibitem [{\citenamefont {Owerre}(2016)}]{Owerre2016}%
  \BibitemOpen
  \bibfield  {author} {\bibinfo {author} {\bibfnamefont {S.~A.}\ \bibnamefont
  {Owerre}},\ }\href@noop {} {\bibfield  {journal} {\bibinfo  {journal}
  {Journal of Physics: Condensed Matter}\ }\textbf {\bibinfo {volume} {28}},\
  \bibinfo {pages} {386001} (\bibinfo {year} {2016})}\BibitemShut {NoStop}%
\bibitem [{\citenamefont {McClarty}(2022)}]{mcclarty22}%
  \BibitemOpen
  \bibfield  {author} {\bibinfo {author} {\bibfnamefont {P.~A.}\ \bibnamefont
  {McClarty}},\ }\href@noop {} {\bibfield  {journal} {\bibinfo  {journal}
  {Annual Review of Condensed Matter Physics}\ }\textbf {\bibinfo {volume}
  {13}},\ \bibinfo {pages} {171} (\bibinfo {year} {2022})}\BibitemShut
  {NoStop}%
\bibitem [{\citenamefont {Bunkov}\ and\ \citenamefont
  {Volovik}(2010)}]{bunkov2010}%
  \BibitemOpen
  \bibfield  {author} {\bibinfo {author} {\bibfnamefont {Y.~M.}\ \bibnamefont
  {Bunkov}}\ and\ \bibinfo {author} {\bibfnamefont {G.~E.}\ \bibnamefont
  {Volovik}},\ }\href@noop {} {\bibfield  {journal} {\bibinfo  {journal}
  {Journal of Physics: Condensed Matter}\ }\textbf {\bibinfo {volume} {22}},\
  \bibinfo {pages} {164210} (\bibinfo {year} {2010})}\BibitemShut {NoStop}%
\bibitem [{\citenamefont {Yuan}\ \emph {et~al.}(2022)\citenamefont {Yuan},
  \citenamefont {Cao}, \citenamefont {Kamra}, \citenamefont {Duine},\ and\
  \citenamefont {Yan}}]{yuan22}%
  \BibitemOpen
  \bibfield  {author} {\bibinfo {author} {\bibfnamefont {H.}~\bibnamefont
  {Yuan}}, \bibinfo {author} {\bibfnamefont {Y.}~\bibnamefont {Cao}}, \bibinfo
  {author} {\bibfnamefont {A.}~\bibnamefont {Kamra}}, \bibinfo {author}
  {\bibfnamefont {R.~A.}\ \bibnamefont {Duine}},\ and\ \bibinfo {author}
  {\bibfnamefont {P.}~\bibnamefont {Yan}},\ }\href@noop {} {\bibfield
  {journal} {\bibinfo  {journal} {Physics Reports}\ }\textbf {\bibinfo {volume}
  {965}},\ \bibinfo {pages} {1} (\bibinfo {year} {2022})}\BibitemShut {NoStop}%
\bibitem [{\citenamefont {Sloan}\ \emph {et~al.}(2019)\citenamefont {Sloan},
  \citenamefont {Rivera}, \citenamefont {Joannopoulos}, \citenamefont
  {Kaminer},\ and\ \citenamefont {Solja\ifmmode \check{c}\else
  \v{c}\fi{}i\ifmmode~\acute{c}\else \'{c}\fi{}}}]{Soljacic2019}%
  \BibitemOpen
  \bibfield  {author} {\bibinfo {author} {\bibfnamefont {J.}~\bibnamefont
  {Sloan}}, \bibinfo {author} {\bibfnamefont {N.}~\bibnamefont {Rivera}},
  \bibinfo {author} {\bibfnamefont {J.~D.}\ \bibnamefont {Joannopoulos}},
  \bibinfo {author} {\bibfnamefont {I.}~\bibnamefont {Kaminer}},\ and\ \bibinfo
  {author} {\bibfnamefont {M.}~\bibnamefont {Solja\ifmmode \check{c}\else
  \v{c}\fi{}i\ifmmode~\acute{c}\else \'{c}\fi{}}},\ }\href@noop {} {\bibfield
  {journal} {\bibinfo  {journal} {Phys. Rev. B}\ }\textbf {\bibinfo {volume}
  {100}},\ \bibinfo {pages} {235453} (\bibinfo {year} {2019})}\BibitemShut
  {NoStop}%
\bibitem [{\citenamefont {Dirnberger}\ \emph {et~al.}(2022)\citenamefont
  {Dirnberger}, \citenamefont {Bushati}, \citenamefont {Datta}, \citenamefont
  {Kumar}, \citenamefont {MacDonald}, \citenamefont {Baldini},\ and\
  \citenamefont {Menon}}]{Dirnberger2022}%
  \BibitemOpen
  \bibfield  {author} {\bibinfo {author} {\bibfnamefont {F.}~\bibnamefont
  {Dirnberger}}, \bibinfo {author} {\bibfnamefont {R.}~\bibnamefont {Bushati}},
  \bibinfo {author} {\bibfnamefont {B.}~\bibnamefont {Datta}}, \bibinfo
  {author} {\bibfnamefont {A.}~\bibnamefont {Kumar}}, \bibinfo {author}
  {\bibfnamefont {A.~H.}\ \bibnamefont {MacDonald}}, \bibinfo {author}
  {\bibfnamefont {E.}~\bibnamefont {Baldini}},\ and\ \bibinfo {author}
  {\bibfnamefont {V.~M.}\ \bibnamefont {Menon}},\ }\href@noop {} {\bibfield
  {journal} {\bibinfo  {journal} {Nature Nanotechnology}\ }\textbf {\bibinfo
  {volume} {17}},\ \bibinfo {pages} {1060} (\bibinfo {year}
  {2022})}\BibitemShut {NoStop}%
\bibitem [{\citenamefont {Mermin}\ and\ \citenamefont
  {Wagner}(1966)}]{mermin66}%
  \BibitemOpen
  \bibfield  {author} {\bibinfo {author} {\bibfnamefont {N.~D.}\ \bibnamefont
  {Mermin}}\ and\ \bibinfo {author} {\bibfnamefont {H.}~\bibnamefont
  {Wagner}},\ }\href {https://doi.org/10.1103/PhysRevLett.17.1133} {\bibfield
  {journal} {\bibinfo  {journal} {\prl}\ }\textbf {\bibinfo {volume} {17}},\
  \bibinfo {pages} {1133} (\bibinfo {year} {1966})}\BibitemShut {NoStop}%
\bibitem [{\citenamefont {Lyu}\ \emph {et~al.}(2020)\citenamefont {Lyu},
  \citenamefont {Gao}, \citenamefont {Zhang}, \citenamefont {Wang},
  \citenamefont {Wu}, \citenamefont {Chen}, \citenamefont {Zhang},
  \citenamefont {Li}, \citenamefont {Huang}, \citenamefont {Zhang},
  \citenamefont {Chen}, \citenamefont {Mei}, \citenamefont {Yan}, \citenamefont
  {Zhao}, \citenamefont {Huang},\ and\ \citenamefont {Huang}}]{VI3_Raman}%
  \BibitemOpen
  \bibfield  {author} {\bibinfo {author} {\bibfnamefont {B.}~\bibnamefont
  {Lyu}}, \bibinfo {author} {\bibfnamefont {Y.}~\bibnamefont {Gao}}, \bibinfo
  {author} {\bibfnamefont {Y.}~\bibnamefont {Zhang}}, \bibinfo {author}
  {\bibfnamefont {L.}~\bibnamefont {Wang}}, \bibinfo {author} {\bibfnamefont
  {X.}~\bibnamefont {Wu}}, \bibinfo {author} {\bibfnamefont {Y.}~\bibnamefont
  {Chen}}, \bibinfo {author} {\bibfnamefont {J.}~\bibnamefont {Zhang}},
  \bibinfo {author} {\bibfnamefont {G.}~\bibnamefont {Li}}, \bibinfo {author}
  {\bibfnamefont {Q.}~\bibnamefont {Huang}}, \bibinfo {author} {\bibfnamefont
  {N.}~\bibnamefont {Zhang}}, \bibinfo {author} {\bibfnamefont
  {Y.}~\bibnamefont {Chen}}, \bibinfo {author} {\bibfnamefont {J.}~\bibnamefont
  {Mei}}, \bibinfo {author} {\bibfnamefont {H.}~\bibnamefont {Yan}}, \bibinfo
  {author} {\bibfnamefont {Y.}~\bibnamefont {Zhao}}, \bibinfo {author}
  {\bibfnamefont {L.}~\bibnamefont {Huang}},\ and\ \bibinfo {author}
  {\bibfnamefont {M.}~\bibnamefont {Huang}},\ }\href@noop {} {\bibfield
  {journal} {\bibinfo  {journal} {Nano Letters}\ }\textbf {\bibinfo {volume}
  {20}},\ \bibinfo {pages} {6024} (\bibinfo {year} {2020})}\BibitemShut
  {NoStop}%
\bibitem [{\citenamefont {Chen}\ \emph {et~al.}(2018)\citenamefont {Chen},
  \citenamefont {Chung}, \citenamefont {Gao}, \citenamefont {Chen},
  \citenamefont {Stone}, \citenamefont {Kolesnikov}, \citenamefont {Huang},\
  and\ \citenamefont {Dai}}]{CrI3neutron2018}%
  \BibitemOpen
  \bibfield  {author} {\bibinfo {author} {\bibfnamefont {L.}~\bibnamefont
  {Chen}}, \bibinfo {author} {\bibfnamefont {J.-H.}\ \bibnamefont {Chung}},
  \bibinfo {author} {\bibfnamefont {B.}~\bibnamefont {Gao}}, \bibinfo {author}
  {\bibfnamefont {T.}~\bibnamefont {Chen}}, \bibinfo {author} {\bibfnamefont
  {M.~B.}\ \bibnamefont {Stone}}, \bibinfo {author} {\bibfnamefont {A.~I.}\
  \bibnamefont {Kolesnikov}}, \bibinfo {author} {\bibfnamefont
  {Q.}~\bibnamefont {Huang}},\ and\ \bibinfo {author} {\bibfnamefont
  {P.}~\bibnamefont {Dai}},\ }\href@noop {} {\bibfield  {journal} {\bibinfo
  {journal} {Phys. Rev. X}\ }\textbf {\bibinfo {volume} {8}},\ \bibinfo {pages}
  {041028} (\bibinfo {year} {2018})}\BibitemShut {NoStop}%
\bibitem [{\citenamefont {Klein}\ \emph {et~al.}(2008)\citenamefont {Klein},
  \citenamefont {de~Loubens}, \citenamefont {Naletov}, \citenamefont {Boust},
  \citenamefont {Guillet}, \citenamefont {Hurdequint}, \citenamefont
  {Leksikov}, \citenamefont {Slavin}, \citenamefont {Tiberkevich},\ and\
  \citenamefont {Vukadinovic}}]{PhysRevB.78.144410}%
  \BibitemOpen
  \bibfield  {author} {\bibinfo {author} {\bibfnamefont {O.}~\bibnamefont
  {Klein}}, \bibinfo {author} {\bibfnamefont {G.}~\bibnamefont {de~Loubens}},
  \bibinfo {author} {\bibfnamefont {V.~V.}\ \bibnamefont {Naletov}}, \bibinfo
  {author} {\bibfnamefont {F.}~\bibnamefont {Boust}}, \bibinfo {author}
  {\bibfnamefont {T.}~\bibnamefont {Guillet}}, \bibinfo {author} {\bibfnamefont
  {H.}~\bibnamefont {Hurdequint}}, \bibinfo {author} {\bibfnamefont
  {A.}~\bibnamefont {Leksikov}}, \bibinfo {author} {\bibfnamefont {A.~N.}\
  \bibnamefont {Slavin}}, \bibinfo {author} {\bibfnamefont {V.~S.}\
  \bibnamefont {Tiberkevich}},\ and\ \bibinfo {author} {\bibfnamefont
  {N.}~\bibnamefont {Vukadinovic}},\ }\href@noop {} {\bibfield  {journal}
  {\bibinfo  {journal} {Phys. Rev. B}\ }\textbf {\bibinfo {volume} {78}},\
  \bibinfo {pages} {144410} (\bibinfo {year} {2008})}\BibitemShut {NoStop}%
\bibitem [{\citenamefont {{Zare Rameshti}}\ \emph {et~al.}(2022)\citenamefont
  {{Zare Rameshti}}, \citenamefont {{Viola Kusminskiy}}, \citenamefont {Haigh},
  \citenamefont {Usami}, \citenamefont {Lachance-Quirion}, \citenamefont
  {Nakamura}, \citenamefont {Hu}, \citenamefont {Tang}, \citenamefont {Bauer},\
  and\ \citenamefont {Blanter}}]{CavityMagnonicsBlanter2021}%
  \BibitemOpen
  \bibfield  {author} {\bibinfo {author} {\bibfnamefont {B.}~\bibnamefont
  {{Zare Rameshti}}}, \bibinfo {author} {\bibfnamefont {S.}~\bibnamefont
  {{Viola Kusminskiy}}}, \bibinfo {author} {\bibfnamefont {J.~A.}\ \bibnamefont
  {Haigh}}, \bibinfo {author} {\bibfnamefont {K.}~\bibnamefont {Usami}},
  \bibinfo {author} {\bibfnamefont {D.}~\bibnamefont {Lachance-Quirion}},
  \bibinfo {author} {\bibfnamefont {Y.}~\bibnamefont {Nakamura}}, \bibinfo
  {author} {\bibfnamefont {C.-M.}\ \bibnamefont {Hu}}, \bibinfo {author}
  {\bibfnamefont {H.~X.}\ \bibnamefont {Tang}}, \bibinfo {author}
  {\bibfnamefont {G.~E.}\ \bibnamefont {Bauer}},\ and\ \bibinfo {author}
  {\bibfnamefont {Y.~M.}\ \bibnamefont {Blanter}},\ }\href
  {https://doi.org/https://doi.org/10.1016/j.physrep.2022.06.001} {\bibfield
  {journal} {\bibinfo  {journal} {Physics Reports}\ }\textbf {\bibinfo {volume}
  {979}},\ \bibinfo {pages} {1} (\bibinfo {year} {2022})},\ \bibinfo {note}
  {cavity Magnonics}\BibitemShut {NoStop}%
\bibitem [{\citenamefont {Tabuchi}\ \emph {et~al.}(2014)\citenamefont
  {Tabuchi}, \citenamefont {Ishino}, \citenamefont {Ishikawa}, \citenamefont
  {Yamazaki}, \citenamefont {Usami},\ and\ \citenamefont
  {Nakamura}}]{PhysRevLett.113.083603}%
  \BibitemOpen
  \bibfield  {author} {\bibinfo {author} {\bibfnamefont {Y.}~\bibnamefont
  {Tabuchi}}, \bibinfo {author} {\bibfnamefont {S.}~\bibnamefont {Ishino}},
  \bibinfo {author} {\bibfnamefont {T.}~\bibnamefont {Ishikawa}}, \bibinfo
  {author} {\bibfnamefont {R.}~\bibnamefont {Yamazaki}}, \bibinfo {author}
  {\bibfnamefont {K.}~\bibnamefont {Usami}},\ and\ \bibinfo {author}
  {\bibfnamefont {Y.}~\bibnamefont {Nakamura}},\ }\href@noop {} {\bibfield
  {journal} {\bibinfo  {journal} {Phys. Rev. Lett.}\ }\textbf {\bibinfo
  {volume} {113}},\ \bibinfo {pages} {083603} (\bibinfo {year}
  {2014})}\BibitemShut {NoStop}%
\bibitem [{\citenamefont {Flower}\ \emph {et~al.}(2019)\citenamefont {Flower},
  \citenamefont {Goryachev}, \citenamefont {Bourhill},\ and\ \citenamefont
  {Tobar}}]{Flower_2019}%
  \BibitemOpen
  \bibfield  {author} {\bibinfo {author} {\bibfnamefont {G.}~\bibnamefont
  {Flower}}, \bibinfo {author} {\bibfnamefont {M.}~\bibnamefont {Goryachev}},
  \bibinfo {author} {\bibfnamefont {J.}~\bibnamefont {Bourhill}},\ and\
  \bibinfo {author} {\bibfnamefont {M.~E.}\ \bibnamefont {Tobar}},\ }\href@noop
  {} {\bibfield  {journal} {\bibinfo  {journal} {New Journal of Physics}\
  }\textbf {\bibinfo {volume} {21}},\ \bibinfo {pages} {095004} (\bibinfo
  {year} {2019})}\BibitemShut {NoStop}%
\bibitem [{\citenamefont {Golovchanskiy}\ \emph {et~al.}(2021)\citenamefont
  {Golovchanskiy}, \citenamefont {Abramov}, \citenamefont {Stolyarov},
  \citenamefont {Golubov}, \citenamefont {Kupriyanov}, \citenamefont
  {Ryazanov},\ and\ \citenamefont {Ustinov}}]{Golovchanskiy2021}%
  \BibitemOpen
  \bibfield  {author} {\bibinfo {author} {\bibfnamefont {I.}~\bibnamefont
  {Golovchanskiy}}, \bibinfo {author} {\bibfnamefont {N.}~\bibnamefont
  {Abramov}}, \bibinfo {author} {\bibfnamefont {V.}~\bibnamefont {Stolyarov}},
  \bibinfo {author} {\bibfnamefont {A.}~\bibnamefont {Golubov}}, \bibinfo
  {author} {\bibfnamefont {M.~Y.}\ \bibnamefont {Kupriyanov}}, \bibinfo
  {author} {\bibfnamefont {V.}~\bibnamefont {Ryazanov}},\ and\ \bibinfo
  {author} {\bibfnamefont {A.}~\bibnamefont {Ustinov}},\ }\href
  {https://doi.org/10.1103/PhysRevApplied.16.034029} {\bibfield  {journal}
  {\bibinfo  {journal} {Phys. Rev. Applied}\ }\textbf {\bibinfo {volume}
  {16}},\ \bibinfo {pages} {034029} (\bibinfo {year} {2021})}\BibitemShut
  {NoStop}%
\bibitem [{\citenamefont {Respaud}\ \emph {et~al.}(1999)\citenamefont
  {Respaud}, \citenamefont {Goiran}, \citenamefont {Broto}, \citenamefont
  {Yang}, \citenamefont {Ould~Ely}, \citenamefont {Amiens},\ and\ \citenamefont
  {Chaudret}}]{HFFMR1999}%
  \BibitemOpen
  \bibfield  {author} {\bibinfo {author} {\bibfnamefont {M.}~\bibnamefont
  {Respaud}}, \bibinfo {author} {\bibfnamefont {M.}~\bibnamefont {Goiran}},
  \bibinfo {author} {\bibfnamefont {J.~M.}\ \bibnamefont {Broto}}, \bibinfo
  {author} {\bibfnamefont {F.~H.}\ \bibnamefont {Yang}}, \bibinfo {author}
  {\bibfnamefont {T.}~\bibnamefont {Ould~Ely}}, \bibinfo {author}
  {\bibfnamefont {C.}~\bibnamefont {Amiens}},\ and\ \bibinfo {author}
  {\bibfnamefont {B.}~\bibnamefont {Chaudret}},\ }\href@noop {} {\bibfield
  {journal} {\bibinfo  {journal} {Phys. Rev. B}\ }\textbf {\bibinfo {volume}
  {59}},\ \bibinfo {pages} {R3934} (\bibinfo {year} {1999})}\BibitemShut
  {NoStop}%
\bibitem [{\citenamefont {Guo}\ \emph {et~al.}()\citenamefont {Guo},
  \citenamefont {Lyu}, \citenamefont {Chen}, \citenamefont {Luo}, \citenamefont
  {Wu}, \citenamefont {Yang}, \citenamefont {Sun}, \citenamefont {de~Abajo},
  \citenamefont {Yang},\ and\ \citenamefont {Dai}}]{GarciadeAbajo2022}%
  \BibitemOpen
  \bibfield  {author} {\bibinfo {author} {\bibfnamefont {X.}~\bibnamefont
  {Guo}}, \bibinfo {author} {\bibfnamefont {W.}~\bibnamefont {Lyu}}, \bibinfo
  {author} {\bibfnamefont {T.}~\bibnamefont {Chen}}, \bibinfo {author}
  {\bibfnamefont {Y.}~\bibnamefont {Luo}}, \bibinfo {author} {\bibfnamefont
  {C.}~\bibnamefont {Wu}}, \bibinfo {author} {\bibfnamefont {B.}~\bibnamefont
  {Yang}}, \bibinfo {author} {\bibfnamefont {Z.}~\bibnamefont {Sun}}, \bibinfo
  {author} {\bibfnamefont {F.~J.~G.}\ \bibnamefont {de~Abajo}}, \bibinfo
  {author} {\bibfnamefont {X.}~\bibnamefont {Yang}},\ and\ \bibinfo {author}
  {\bibfnamefont {Q.}~\bibnamefont {Dai}},\ }\href@noop {} {\bibfield
  {journal} {\bibinfo  {journal} {Advanced Materials}\ }\textbf {\bibinfo
  {volume} {n/a}},\ \bibinfo {pages} {2201856}}\BibitemShut {NoStop}%
\bibitem [{\citenamefont {Costa}\ \emph {et~al.}(2021)\citenamefont {Costa},
  \citenamefont {Gonçalves}, \citenamefont {Basov}, \citenamefont {Koppens},
  \citenamefont {Mortensen},\ and\ \citenamefont {Peres}}]{Costa2021PNAS}%
  \BibitemOpen
  \bibfield  {author} {\bibinfo {author} {\bibfnamefont {A.~T.}\ \bibnamefont
  {Costa}}, \bibinfo {author} {\bibfnamefont {P.~A.~D.}\ \bibnamefont
  {Gonçalves}}, \bibinfo {author} {\bibfnamefont {D.~N.}\ \bibnamefont
  {Basov}}, \bibinfo {author} {\bibfnamefont {F.~H.~L.}\ \bibnamefont
  {Koppens}}, \bibinfo {author} {\bibfnamefont {N.~A.}\ \bibnamefont
  {Mortensen}},\ and\ \bibinfo {author} {\bibfnamefont {N.~M.~R.}\ \bibnamefont
  {Peres}},\ }\href@noop {} {\bibfield  {journal} {\bibinfo  {journal}
  {Proceedings of the National Academy of Sciences}\ }\textbf {\bibinfo
  {volume} {118}},\ \bibinfo {pages} {e2012847118} (\bibinfo {year}
  {2021})}\BibitemShut {NoStop}%
\bibitem [{\citenamefont {Costa}\ and\ \citenamefont
  {Peres}(2021)}]{Costa2021JPCM}%
  \BibitemOpen
  \bibfield  {author} {\bibinfo {author} {\bibfnamefont {A.~T.}\ \bibnamefont
  {Costa}}\ and\ \bibinfo {author} {\bibfnamefont {N.~M.~R.}\ \bibnamefont
  {Peres}},\ }\href@noop {} {\bibfield  {journal} {\bibinfo  {journal} {Journal
  of Physics: Condensed Matter}\ }\textbf {\bibinfo {volume} {34}},\ \bibinfo
  {pages} {105304} (\bibinfo {year} {2021})}\BibitemShut {NoStop}%
\bibitem [{\citenamefont {Nerl}\ \emph {et~al.}(2017)\citenamefont {Nerl},
  \citenamefont {Winther}, \citenamefont {Hage}, \citenamefont {Thygesen},
  \citenamefont {Houben}, \citenamefont {Backes}, \citenamefont {Coleman},
  \citenamefont {Ramasse},\ and\ \citenamefont {Nicolosi}}]{Nerl2017}%
  \BibitemOpen
  \bibfield  {author} {\bibinfo {author} {\bibfnamefont {H.~C.}\ \bibnamefont
  {Nerl}}, \bibinfo {author} {\bibfnamefont {K.~T.}\ \bibnamefont {Winther}},
  \bibinfo {author} {\bibfnamefont {F.~S.}\ \bibnamefont {Hage}}, \bibinfo
  {author} {\bibfnamefont {K.~S.}\ \bibnamefont {Thygesen}}, \bibinfo {author}
  {\bibfnamefont {L.}~\bibnamefont {Houben}}, \bibinfo {author} {\bibfnamefont
  {C.}~\bibnamefont {Backes}}, \bibinfo {author} {\bibfnamefont {J.~N.}\
  \bibnamefont {Coleman}}, \bibinfo {author} {\bibfnamefont {Q.~M.}\
  \bibnamefont {Ramasse}},\ and\ \bibinfo {author} {\bibfnamefont
  {V.}~\bibnamefont {Nicolosi}},\ }\href@noop {} {\bibfield  {journal}
  {\bibinfo  {journal} {npj 2D Materials and Applications}\ }\textbf {\bibinfo
  {volume} {1}},\ \bibinfo {pages} {2} (\bibinfo {year} {2017})}\BibitemShut
  {NoStop}%
\bibitem [{\citenamefont {Dicke}(1954)}]{Dicke1954}%
  \BibitemOpen
  \bibfield  {author} {\bibinfo {author} {\bibfnamefont {R.~H.}\ \bibnamefont
  {Dicke}},\ }\href@noop {} {\bibfield  {journal} {\bibinfo  {journal} {Phys.
  Rev.}\ }\textbf {\bibinfo {volume} {93}},\ \bibinfo {pages} {99} (\bibinfo
  {year} {1954})}\BibitemShut {NoStop}%
\bibitem [{\citenamefont {Halperin}(2019)}]{Halperin2019}%
  \BibitemOpen
  \bibfield  {author} {\bibinfo {author} {\bibfnamefont {B.~I.}\ \bibnamefont
  {Halperin}},\ }\href@noop {} {\bibfield  {journal} {\bibinfo  {journal}
  {Journal of Statistical Physics}\ }\textbf {\bibinfo {volume} {175}},\
  \bibinfo {pages} {521} (\bibinfo {year} {2019})}\BibitemShut {NoStop}%
\bibitem [{\citenamefont {Lee}\ \emph {et~al.}(2020)\citenamefont {Lee},
  \citenamefont {Utermohlen}, \citenamefont {Weber}, \citenamefont {Hwang},
  \citenamefont {Zhang}, \citenamefont {van Tol}, \citenamefont {Goldberger},
  \citenamefont {Trivedi},\ and\ \citenamefont {Hammel}}]{FMR_CrI3PRL2019}%
  \BibitemOpen
  \bibfield  {author} {\bibinfo {author} {\bibfnamefont {I.}~\bibnamefont
  {Lee}}, \bibinfo {author} {\bibfnamefont {F.~G.}\ \bibnamefont {Utermohlen}},
  \bibinfo {author} {\bibfnamefont {D.}~\bibnamefont {Weber}}, \bibinfo
  {author} {\bibfnamefont {K.}~\bibnamefont {Hwang}}, \bibinfo {author}
  {\bibfnamefont {C.}~\bibnamefont {Zhang}}, \bibinfo {author} {\bibfnamefont
  {J.}~\bibnamefont {van Tol}}, \bibinfo {author} {\bibfnamefont {J.~E.}\
  \bibnamefont {Goldberger}}, \bibinfo {author} {\bibfnamefont
  {N.}~\bibnamefont {Trivedi}},\ and\ \bibinfo {author} {\bibfnamefont {P.~C.}\
  \bibnamefont {Hammel}},\ }\href@noop {} {\bibfield  {journal} {\bibinfo
  {journal} {Phys. Rev. Lett.}\ }\textbf {\bibinfo {volume} {124}},\ \bibinfo
  {pages} {017201} (\bibinfo {year} {2020})}\BibitemShut {NoStop}%
\bibitem [{\citenamefont {You}\ \emph {et~al.}(2019)\citenamefont {You},
  \citenamefont {Chen}, \citenamefont {Zhang}, \citenamefont {Sheng},
  \citenamefont {Yang},\ and\ \citenamefont {Su}}]{PtCl32019}%
  \BibitemOpen
  \bibfield  {author} {\bibinfo {author} {\bibfnamefont {J.-Y.}\ \bibnamefont
  {You}}, \bibinfo {author} {\bibfnamefont {C.}~\bibnamefont {Chen}}, \bibinfo
  {author} {\bibfnamefont {Z.}~\bibnamefont {Zhang}}, \bibinfo {author}
  {\bibfnamefont {X.-L.}\ \bibnamefont {Sheng}}, \bibinfo {author}
  {\bibfnamefont {S.~A.}\ \bibnamefont {Yang}},\ and\ \bibinfo {author}
  {\bibfnamefont {G.}~\bibnamefont {Su}},\ }\href@noop {} {\bibfield  {journal}
  {\bibinfo  {journal} {Phys. Rev. B}\ }\textbf {\bibinfo {volume} {100}},\
  \bibinfo {pages} {064408} (\bibinfo {year} {2019})}\BibitemShut {NoStop}%
\bibitem [{\citenamefont {Jiang}\ \emph {et~al.}(2021)\citenamefont {Jiang},
  \citenamefont {Liu}, \citenamefont {Xing}, \citenamefont {Liu}, \citenamefont
  {Guo}, \citenamefont {Liu},\ and\ \citenamefont {Zhao}}]{Jiang2021}%
  \BibitemOpen
  \bibfield  {author} {\bibinfo {author} {\bibfnamefont {X.}~\bibnamefont
  {Jiang}}, \bibinfo {author} {\bibfnamefont {Q.}~\bibnamefont {Liu}}, \bibinfo
  {author} {\bibfnamefont {J.}~\bibnamefont {Xing}}, \bibinfo {author}
  {\bibfnamefont {N.}~\bibnamefont {Liu}}, \bibinfo {author} {\bibfnamefont
  {Y.}~\bibnamefont {Guo}}, \bibinfo {author} {\bibfnamefont {Z.}~\bibnamefont
  {Liu}},\ and\ \bibinfo {author} {\bibfnamefont {J.}~\bibnamefont {Zhao}},\
  }\href@noop {} {\bibfield  {journal} {\bibinfo  {journal} {Applied Physics
  Reviews}\ }\textbf {\bibinfo {volume} {8}},\ \bibinfo {pages} {031305}
  (\bibinfo {year} {2021})}\BibitemShut {NoStop}%
\bibitem [{\citenamefont {Yen}\ \emph {et~al.}(2004)\citenamefont {Yen},
  \citenamefont {Padilla}, \citenamefont {Fang}, \citenamefont {Vier},
  \citenamefont {Smith}, \citenamefont {Pendry}, \citenamefont {Basov},\ and\
  \citenamefont {Zhang}}]{Pendry2004}%
  \BibitemOpen
  \bibfield  {author} {\bibinfo {author} {\bibfnamefont {T.~J.}\ \bibnamefont
  {Yen}}, \bibinfo {author} {\bibfnamefont {W.~J.}\ \bibnamefont {Padilla}},
  \bibinfo {author} {\bibfnamefont {N.}~\bibnamefont {Fang}}, \bibinfo {author}
  {\bibfnamefont {D.~C.}\ \bibnamefont {Vier}}, \bibinfo {author}
  {\bibfnamefont {D.~R.}\ \bibnamefont {Smith}}, \bibinfo {author}
  {\bibfnamefont {J.~B.}\ \bibnamefont {Pendry}}, \bibinfo {author}
  {\bibfnamefont {D.~N.}\ \bibnamefont {Basov}},\ and\ \bibinfo {author}
  {\bibfnamefont {X.}~\bibnamefont {Zhang}},\ }\href@noop {} {\bibfield
  {journal} {\bibinfo  {journal} {Science}\ }\textbf {\bibinfo {volume}
  {303}},\ \bibinfo {pages} {1494} (\bibinfo {year} {2004})}\BibitemShut
  {NoStop}%
\bibitem [{\citenamefont {Lado}\ and\ \citenamefont
  {Fern{\'a}ndez-Rossier}(2017)}]{lado2017}%
  \BibitemOpen
  \bibfield  {author} {\bibinfo {author} {\bibfnamefont {J.~L.}\ \bibnamefont
  {Lado}}\ and\ \bibinfo {author} {\bibfnamefont {J.}~\bibnamefont
  {Fern{\'a}ndez-Rossier}},\ }\href@noop {} {\bibfield  {journal} {\bibinfo
  {journal} {2D Materials}\ }\textbf {\bibinfo {volume} {4}},\ \bibinfo {pages}
  {035002} (\bibinfo {year} {2017})}\BibitemShut {NoStop}%
\bibitem [{\citenamefont {Henriques}\ \emph {et~al.}(2021)\citenamefont
  {Henriques}, \citenamefont {Amorim},\ and\ \citenamefont
  {Peres}}]{Henriques2021}%
  \BibitemOpen
  \bibfield  {author} {\bibinfo {author} {\bibfnamefont {J.~C.~G.}\
  \bibnamefont {Henriques}}, \bibinfo {author} {\bibfnamefont {B.}~\bibnamefont
  {Amorim}},\ and\ \bibinfo {author} {\bibfnamefont {N.~M.~R.}\ \bibnamefont
  {Peres}},\ }\href@noop {} {\bibfield  {journal} {\bibinfo  {journal} {Phys.
  Rev. B}\ }\textbf {\bibinfo {volume} {103}},\ \bibinfo {pages} {085407}
  (\bibinfo {year} {2021})}\BibitemShut {NoStop}%
\bibitem [{\citenamefont {Gon\c{c}alves}\ and\ \citenamefont
  {Peres}(2016)}]{NunosBook}%
  \BibitemOpen
  \bibfield  {author} {\bibinfo {author} {\bibfnamefont {P.~A.~D.}\
  \bibnamefont {Gon\c{c}alves}}\ and\ \bibinfo {author} {\bibfnamefont
  {N.~M.~R.}\ \bibnamefont {Peres}},\ }\href@noop {} {\emph {\bibinfo {title}
  {An Introduction to Graphene Plasmonics}}},\ \bibinfo {edition} {1st}\ ed.\
  (\bibinfo  {publisher} {World Scientific},\ \bibinfo {address} {Singapore},\
  \bibinfo {year} {2016})\BibitemShut {NoStop}%
\bibitem [{\citenamefont {Holstein}\ and\ \citenamefont
  {Primakoff}(1940)}]{holstein40}%
  \BibitemOpen
  \bibfield  {author} {\bibinfo {author} {\bibfnamefont {T.}~\bibnamefont
  {Holstein}}\ and\ \bibinfo {author} {\bibfnamefont {H.}~\bibnamefont
  {Primakoff}},\ }\href@noop {} {\bibfield  {journal} {\bibinfo  {journal}
  {Physical Review}\ }\textbf {\bibinfo {volume} {58}},\ \bibinfo {pages}
  {1098} (\bibinfo {year} {1940})}\BibitemShut {NoStop}%
\bibitem [{\citenamefont {Calder}\ \emph {et~al.}(2019)\citenamefont {Calder},
  \citenamefont {Kolesnikov},\ and\ \citenamefont {May}}]{FeGeTe_neutron}%
  \BibitemOpen
  \bibfield  {author} {\bibinfo {author} {\bibfnamefont {S.}~\bibnamefont
  {Calder}}, \bibinfo {author} {\bibfnamefont {A.~I.}\ \bibnamefont
  {Kolesnikov}},\ and\ \bibinfo {author} {\bibfnamefont {A.~F.}\ \bibnamefont
  {May}},\ }\href@noop {} {\bibfield  {journal} {\bibinfo  {journal} {Phys.
  Rev. B}\ }\textbf {\bibinfo {volume} {99}},\ \bibinfo {pages} {094423}
  (\bibinfo {year} {2019})}\BibitemShut {NoStop}%
\bibitem [{\citenamefont {Park}\ \emph {et~al.}(2020)\citenamefont {Park},
  \citenamefont {Kim}, \citenamefont {Liu}, \citenamefont {Hwang},
  \citenamefont {Kim}, \citenamefont {Kim}, \citenamefont {Kim}, \citenamefont
  {Petrovic}, \citenamefont {Hwang}, \citenamefont {Mo}, \citenamefont {Kim},
  \citenamefont {Min}, \citenamefont {Koo}, \citenamefont {Chang},
  \citenamefont {Jang}, \citenamefont {Choi},\ and\ \citenamefont
  {Ryu}}]{FeGeTeMAE2020}%
  \BibitemOpen
  \bibfield  {author} {\bibinfo {author} {\bibfnamefont {S.~Y.}\ \bibnamefont
  {Park}}, \bibinfo {author} {\bibfnamefont {D.~S.}\ \bibnamefont {Kim}},
  \bibinfo {author} {\bibfnamefont {Y.}~\bibnamefont {Liu}}, \bibinfo {author}
  {\bibfnamefont {J.}~\bibnamefont {Hwang}}, \bibinfo {author} {\bibfnamefont
  {Y.}~\bibnamefont {Kim}}, \bibinfo {author} {\bibfnamefont {W.}~\bibnamefont
  {Kim}}, \bibinfo {author} {\bibfnamefont {J.-Y.}\ \bibnamefont {Kim}},
  \bibinfo {author} {\bibfnamefont {C.}~\bibnamefont {Petrovic}}, \bibinfo
  {author} {\bibfnamefont {C.}~\bibnamefont {Hwang}}, \bibinfo {author}
  {\bibfnamefont {S.-K.}\ \bibnamefont {Mo}}, \bibinfo {author} {\bibfnamefont
  {H.-j.}\ \bibnamefont {Kim}}, \bibinfo {author} {\bibfnamefont {B.-C.}\
  \bibnamefont {Min}}, \bibinfo {author} {\bibfnamefont {H.~C.}\ \bibnamefont
  {Koo}}, \bibinfo {author} {\bibfnamefont {J.}~\bibnamefont {Chang}}, \bibinfo
  {author} {\bibfnamefont {C.}~\bibnamefont {Jang}}, \bibinfo {author}
  {\bibfnamefont {J.~W.}\ \bibnamefont {Choi}},\ and\ \bibinfo {author}
  {\bibfnamefont {H.}~\bibnamefont {Ryu}},\ }\href@noop {} {\bibfield
  {journal} {\bibinfo  {journal} {Nano Letters}\ }\textbf {\bibinfo {volume}
  {20}},\ \bibinfo {pages} {95} (\bibinfo {year} {2020})}\BibitemShut {NoStop}%
\bibitem [{\citenamefont {Principi}\ \emph {et~al.}(2013)\citenamefont
  {Principi}, \citenamefont {Vignale}, \citenamefont {Carrega},\ and\
  \citenamefont {Polini}}]{Principi2013}%
  \BibitemOpen
  \bibfield  {author} {\bibinfo {author} {\bibfnamefont {A.}~\bibnamefont
  {Principi}}, \bibinfo {author} {\bibfnamefont {G.}~\bibnamefont {Vignale}},
  \bibinfo {author} {\bibfnamefont {M.}~\bibnamefont {Carrega}},\ and\ \bibinfo
  {author} {\bibfnamefont {M.}~\bibnamefont {Polini}},\ }\href@noop {}
  {\bibfield  {journal} {\bibinfo  {journal} {Phys. Rev. B}\ }\textbf {\bibinfo
  {volume} {88}},\ \bibinfo {pages} {195405} (\bibinfo {year}
  {2013})}\BibitemShut {NoStop}%
\bibitem [{Note1()}]{Note1}%
  \BibitemOpen
  \bibinfo {note} {\protect {Other methods for exciting surface
  plasmon-polaritons are available, such as patterning gratings on top of
  graphene, which allow attaining larger wave vectors; here, however, we are
  interested in small wave vectors, for which the ATR is
  appropriate.}}\BibitemShut {Stop}%
\end{thebibliography}%

\onecolumngrid

\appendix

\section{Quantization of the plasmon's electromagnetic field}
The plasmon's magnetic field operator $\hat{\vec{B}}$ at a point $(\vec{r},z)$ is
\begin{equation}
    \hat{\vec{B}}(\vec{r},z)=\sum_{\vec{q}}\left[ \vec{B}_{\vec{q}}(\vec{r},z)a^\dagger_{\vec{q}} + \vec{B}^*_{\vec{q}}(\vec{r},z)a_{\vec{q}}\right],
\end{equation}
where $a^\dagger_{\vec{q}},a_{\vec{q}}$ are the creation and annihilation operators for the plasmon's photonic component with wave vector $\vec{q}$, and the coefficients $\vec{B}_{\vec{q}}(\vec{r},z)$ can be derived from those appearing in the operator for the vector potential,~\cite{Henriques2021}
\begin{equation}
\vec{\mathrm{A}}_{\vec{q}}(\vec{r},z) = \sqrt{\frac{\hbar}{2A\epsilon_0\omega_{\mathrm{pl}}(q)\Lambda(\vec{q})}}
e^{i\vec{q}\cdot\vec{r}}\times\left\{\begin{array}{c} 
\left( i\frac{\vec{q}}{q}-\frac{q}{\kappa_{1,\vec{q}}}\hat{z}\right)e^{-\kappa_{1,\vec{q}}z},\,\, z>0, \\
\left( i\frac{\vec{q}}{q}+\frac{q}{\kappa_{2,\vec{q}}}\hat{z}\right)e^{\kappa_{2,\vec{q}}z},\,\, z<0.    
\end{array}
\right.
\end{equation}
$A$ is the area of the graphene sheet, which is assumed to occupy the plane $z_0$. The dispersion relation for the graphene plasmon, $\omega_{\mathrm{pl}}(q)$, can be found in the main text (eq.~7). The mode length $\Lambda(q)$ is defined as
\begin{eqnarray}
    \Lambda(\vec{q}) = \frac{[\omega_{\mathrm{pl}}(q)c]^2}{2}\left(\frac{\epsilon_1^2}{\kappa^3_{1,\vec{q}}} + 
    \frac{\epsilon_2^2}{\kappa^3_{2,\vec{q}}}\right)+
    \frac{i}{2\epsilon_0}\left[\frac{\sigma(\omega_{\mathrm{pl}}(q))}{\omega_{\mathrm{pl}}(q)} + \left.\frac{\partial\sigma}{\partial\omega}\right|_{\omega_{\mathrm{pl}}(q)}\right],
\end{eqnarray}
where $\sigma(\omega)$ is the conductivity of graphene, $\epsilon_0$ is the permittivity of free space and $\epsilon_j$ are the relative permittivities of the media surrounding graphene. Also, 
\begin{equation}
    \kappa_{j,\vec{q}}\equiv\sqrt{q^2-\epsilon_j\frac{\omega^2_{\mathrm{pl}}(q)}{c^2}},
\end{equation}
The coefficients of the expansion
of the magnetic field operator into normal modes with wave vector $\vec{q}$ is
\begin{equation}
    \vec{B}_{\vec{q}}(\vec{r},z)=\nabla\times\vec{\mathrm{A}}_{\vec{q}}(\vec{r},z) = iF(q,z)e^{i\vec{q}\cdot\vec{r}}\left(q_y\hat{x}-q_x\hat{y}\right),
\end{equation}
with
\begin{eqnarray}
    F(q,z) \equiv 
    -\epsilon_1\frac{\omega^2_{\mathrm{pl}}(q)}{c^2q\kappa_{1,\vec{q}}}\sqrt{\frac{\hbar}{2A\epsilon_0\omega_{\mathrm{pl}}(q)\Lambda(\vec{q})}}
    e^{-\kappa_{1,\vec{q}}z}
\end{eqnarray}
for $z>0$, where we assume the 2D ferromagnet to be placed.

\section{Plasmon Green function}

The two-time, retarded Green function of the plasmon can be defined as  
\begin{equation}
    \gpl(\vec{k},\vec{k}';t)\equiv \gf{a_{\vec{k}}(t);a^\dagger_{\vec{k}'}} \equiv -i\theta(t)\left\langle \left[a_{\vec{k}}(t),a^\dagger_{\vec{k}'}\right]\right\rangle,
\end{equation}
where time evolution is determined in the Heisenberg representation. Wherever the time argument of an operator
is omitted it should be taken as $t=0$.  It is straightforward to show that, since the system is translationally invariant,
$ \gpl(\vec{k},\vec{k}';t)= \gpl(\vec{k};t)\delta_{\vec{k},\vec{k}'}$.
The plasmon Green function obeys the equation of motion,
\begin{equation}
    i\hbar\frac{d}{dt}\gpl(\vec{k},\vec{k}';t) = \delta(t)\left\langle\left[a_{\vec{k}},a^\dagger_{\vec{k}'}\right]\right\rangle + 
    \gf{\left[a_{\vec{k}},H\right](t);a^\dagger_{\vec{k}'}}.
\end{equation}
But,
\begin{equation}
    \left[a_{\vec{k}},H_\mathrm{pl}\right]=\hbar\omega_\mathrm{pl}(k)a_{\vec{k}},
\end{equation}
\begin{equation}
    \left[a_{\vec{k}},H_\mathrm{Z}\right]=\Omega^*_{-\vec{k}}(z)b_{-\vec{k}}+\Omega_{\vec{k}}(z)b^\dagger_{\vec{k}}.
\end{equation}
Thus,
\begin{eqnarray}
    i\hbar\frac{d}{dt}\gpl(\vec{k},\vec{k}';t) = \delta(t)\delta_{\vec{k},\vec{k}'} + 
    \hbar\omega_{\mathrm{pl}}(k)\gpl(\vec{k},\vec{k}';t) +
    \Omega^*_{-\vec{k}}(z)\gf{b_{-\vec{k}}(t);a^\dagger_{\vec{k}'}}+ \Omega_{\vec{k}}(z)\gf{b^\dagger_{\vec{k}}(t);a^\dagger_{\vec{k}'}}.
\end{eqnarray}
Thus, the equation of motion for the plasmon Green function is one of a (closed) system of coupled equations which also involves the magnon Green function,
\begin{equation}
    G(\vec{k},\vec{k}';t)\equiv \gf{b_{\vec{k}}(t);b^\dagger_{\vec{k}'}} \equiv -i\theta(t)\left\langle \left[b_{\vec{k}}(t),b^\dagger_{\vec{k}'}\right]\right\rangle,
\end{equation}
and mixed plasmon-magnon Green functions, such as
\begin{eqnarray}
    \gf{b_{-\vec{k}}(t);a^\dagger_{\vec{k}'}}\equiv -i\theta(t)\left\langle \left[b_{-\vec{k}}(t),a^\dagger_{\vec{k}'}\right]\right\rangle,\\
    \gf{b^\dagger_{\vec{k}}(t);a^\dagger_{\vec{k}'}} \equiv -i\theta(t)\left\langle \left[b^\dagger_{\vec{k}}(t),a^\dagger_{\vec{k}'}\right]\right\rangle.
\end{eqnarray}
After Fourier transforming to the frequency domain and some subsequent algebra, we obtain
\begin{equation}
    \gpl(\vec{k},\vec{k}',\fE) = \frac{\delta_{\vec{k},\vec{k}'}}{\fE - \hbar\omega_\mathrm{pl}(q) -\Sigma_\mathrm{pl}(\vec{k},\fE)},
\end{equation}
where
\begin{eqnarray}
    \Sigma_\mathrm{Pl}(\vec{k},\fE)\equiv |\Omega_{\vec{k}}(z)|^2\left(\frac{1}{\fE-\hbar\omega_{\mathrm{mag}}(\vec{k})}-\frac{1}{\fE+\hbar\omega_{\mathrm{mag}}(-\vec{k})}\right)\times\nonumber\\
    \left[   1 + |\Omega_{\vec{k}}(z)|^2\left(\frac{1}{\fE-\hbar\omega_{\mathrm{mag}}(\vec{k})}-\frac{1}{\fE+\hbar\omega_{\mathrm{mag}}(-\vec{k})}\right)
    \frac{1}{\fE + \hbar\omega_{\mathrm{pl}}(k)}
    \right]^{-1}.
\end{eqnarray}
If the magnon dispersion relation is reciprocal, i.e., $\omega_{\mathrm{mag}}(\vec{k}) = \omega_{\mathrm{mag}}(-\vec{k})$,
\begin{eqnarray}
    \Sigma_\mathrm{pl}(\vec{k},\fE)\equiv 2\hbar\omega_{\mathrm{mag}}(k)|\Omega_{\vec{k}}(z)|^2\left[\fE^2-(\hbar\omega_{\mathrm{mag}}(\vec{k}))^2\right]^{-1}\times\nonumber\\
    \left\{   1 + 2\omega_{\mathrm{mag}}(\vec{k})|\Omega_{\vec{k}}(z)|^2\left[
    {\fE^2-(\hbar\omega_{\mathrm{mag}}(\vec{k}))^2}\right]^{-1}
    \left[\fE + \hbar\omega_{\mathrm{pl}}(k)\right]^{-1}
    \right\}^{-1}.
\end{eqnarray}
We also obtain the magnon Green function,
\begin{eqnarray}
    G(\vec{k},\vec{k}';\fE) = \frac{\delta_{\vec{k},\vec{k}'}}{\fE - \hbar\omega_{\mathrm{mag}}(\vec{k})-\Sigma_{\mathrm{mag}}(\vec{k};\fE)},
\end{eqnarray}
where
\begin{eqnarray}
    \Sigma_{\mathrm{mag}}(\vec{k};\fE)\equiv 
    2\hbar\omega_{\mathrm{pl}}(k)|\Omega_{\vec{k}}(z)|^2\left[\fE^2-(\hbar\omega_{\mathrm{pl}}(k))^2\right]^{-1}\times\nonumber\\
    \left\{ 1+2\hbar\omega_{\mathrm{pl}}(k)|\Omega_{\vec{k}}(z)|^2\left[\fE^2-(\hbar\omega_{\mathrm{pl}}(k))^2\right]^{-1} 
    \left[\fE +\hbar\omega_{\mathrm{mag}}(-\vec{k})\right]^{-1}\right\}^{-1}  .
\end{eqnarray}

\end{document}